\documentclass[twocolumn,showpacs,prl,amsmath,amssymb,superscriptaddress]{revtex4}
\usepackage{graphicx}
\usepackage{dcolumn}
\usepackage{bm}
\usepackage[latin1]{inputenc}
\usepackage[mathscr]{eucal}
\usepackage{epsfig}
\usepackage{rotating}
\usepackage{multirow}

\begin{document}

\title{Recombination limited energy relaxation in a BCS superconductor}


\author{A.V. Timofeev} \affiliation{Low Temperature
Laboratory, Helsinki University of Technology, P.O. Box 3500, 02015
TKK, Finland} \affiliation{Institute of Solid State Physics, Russian
Academy of Sciences, Chernogolovka, 142432 Russia}

\author{C. Pascual García}
\affiliation{NEST CNR-INFM \& Scuola Normale Superiore, I-56126
Pisa, Italy}

\author{N.B. Kopnin}
\affiliation{Low Temperature Laboratory, Helsinki University of
Technology, P.O. Box 3500, 02015 TKK, Finland}
\affiliation{L. D.
Landau Institute for Theoretical Physics, 117940 Moscow, Russia}

\author{A.M. Savin}
\affiliation{Low Temperature Laboratory, Helsinki University of
Technology, P.O. Box 3500, 02015 TKK, Finland}

\author{M. Meschke}
\affiliation{Low Temperature Laboratory, Helsinki University of
Technology, P.O. Box 3500, 02015 TKK, Finland}

\author{F. Giazotto}
\affiliation{NEST CNR-INFM \& Scuola Normale Superiore, I-56126
Pisa, Italy}

\author{J.P. Pekola}
\affiliation{Low Temperature Laboratory,
Helsinki University of Technology, P.O. Box 3500, 02015 TKK,
Finland}

\begin{abstract}
We study quasiparticle energy relaxation at sub-kelvin temperatures
by injecting hot electrons into an aluminium island and measuring
the energy flux from electrons into phonons both in the
superconducting and in the normal state. The data show strong
reduction of the flux at low temperatures in the superconducting
state, in qualitative agreement with the presented quasiclassical
theory for clean superconductors. Quantitatively, the energy flux
exceeds that from the theory both in the superconducting and in the
normal state, possibly suggesting an enhanced or additional
relaxation process.
\end{abstract}


\maketitle

Superconducting nanostructures attract lots of attention currently,
partly because of their potential applications, for instance, in
single Cooper pair and single-electron devices, in quantum
information processing, and in detection of radiation. Although the
operation of many of these devices is based on charge transport, the
energy relaxation in them is also of importance to warrant proper
functioning either under driven conditions, or when subjected to
environment fluctuations. Thermalization of the electron system with
the surrounding bath is a serious concern at sub-kelvin temperatures
for non-superconducting structures, but securing proper
thermalization of a superconductor is an even greater challenge. In
particular, recombination of hot quasiparticles (QP:s) into Cooper
pairs slows down exponentially towards low temperatures. QP
scattering rates in usual BCS superconductors have been assessed
theoretically already several decades ago \cite{kaplan76,reizer89},
and there have been measurements of them, both soon after the first
predictions (for review, see \cite{kaplan76}), and recently also at
very low temperatures \cite{day03,kozorezov01,barends08}. However,
although a well recognized issue in normal systems, the most
relevant property of relaxation, the associated heat flux, has not
been addressed in the past. This is the topic of the present letter.
We present both experimental and theoretical results which
demonstrate the importance of slow thermal relaxation in
superconducting nanostructures.

Energy relaxation in normal metals has been investigated
thoroughly in experiment for a long time
\cite{gantmakher74,roukes85,wellstood94,Girvin96}. The central
results can be summarized as follows. In three dimensional systems
electron-phonon (e-p) heat flux $P_{\rm ep}$ is
\begin{equation} \label{e-p-heat-norm}
P_{\rm ep}=\Sigma \mathcal{V}(T_{\rm e}^5 -T_{\rm p}^5).
\end{equation}
Here, $\Sigma$ is a material constant \cite{giazotto06},
$\mathcal{V}$ is the volume of the electronic system, and $T_{\rm
e}$ and $T_{\rm p}$ are the temperatures of electrons and phonons,
respectively. Deviations from this behaviour towards the fourth
power of temperature have been seen for lower temperatures (see, for
example, Ref. \cite{Girvin96} and also more recent Ref.
\cite{karvonen07}), and are usually explained by the impurity
effects \cite{Altshuler78,ReizerSergeev85,sergeev00} when the wave
length of a thermal phonon becomes of the order of the electron mean
free path or of the sample size. Nevertheless, Eq.
\eqref{e-p-heat-norm} gives a good account of the heat flux observed
in most experiments at sub-kelvin temperatures. Under the same
conditions electron-electron (e-e) relaxation is typically much
faster; most experiments demonstrate the so-called
quasi-equilibrium, where electrons have a well-defined temperature,
usually different from that of the phonons. Deviations from this
simple picture have been observed, e.g., in voltage biased diffusive
wires \cite{pothier97}.

\begin{figure}[t]
\includegraphics[width=0.90\linewidth]{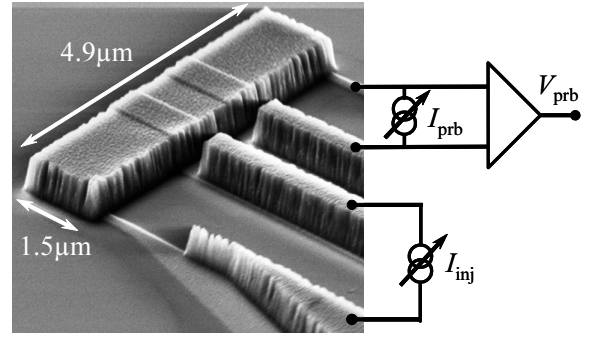}
\caption{A typical sample (Sample C) for measuring energy relaxation
in an aluminium superconducting bar. The circuits on the right
indicate injection of hot QP:s and probing the island temperature.}
\label{sample}
\end{figure}

Relaxation processes in superconductors have also been studied, see,
e.g., Ref. \cite{kaplan76}. The most obvious features different from
the normal state are: (i) The QP:s need to emit or absorb an energy
in excess of the energy gap $\Delta$ to be recombined or excited,
respectively. This leads to exponentially slow e-p relaxation rates
at low temperatures. (ii) The number of QP:s is very small well
below $T_C$, leading to slow e-e relaxation as well. The focus has
been in relaxation times, with no attempts to obtain energy flux in
the spirit of Eq. \eqref{e-p-heat-norm}. The relaxation time was
addressed recently, e.g., in experiments on superconducting photon
detectors \cite{day03,kozorezov01,barends08}; these measurements
suggest to confirm the recombination limited rate $\tau_{\rm
rec}^{-1} \propto \sqrt{T/T_C}e^{-\Delta/k_BT}$ down to $T/T_C
\simeq 0.2$. At lower $T$ the relaxation time saturates due to
presently poorly known reasons.

The e-p processes in clean superconductors can be characterized by
the rate $\tau^{-1}_{\bf k, k-q}$ of a QP with wave vector ${\bf k}$
to emit a phonon with wave vector ${\bf q}$,
$\tau_{\bf k,k-q}^{-1} =  (2\pi/\hbar) |\mathcal{M}_{\bf k,k-q}|^2
\delta(E_{\bf k}-E_{{\bf k-q}}-\epsilon_{\bf q})N(E_{{\bf k-q}})
[1-f(E_{{\bf k-q}})][n_p({\bf q},T_{\rm p})+1]$.
Here, $\mathcal{M}_{\bf k,k-q}$ is the matrix element of
electron-phonon coupling in a superconductor, $f(E)$ is the
distribution function of electrons; it is the Fermi function
$f(E,T_{\rm e}) =(1+e^{E/k_BT_{\rm e}})^{-1}$ if electrons are in
equilibrium with a temperature $T_{\rm e}$. Phonons are assumed to
be in equilibrium with occupation $n_p({\bf q},T_{\rm
p})=(e^{\epsilon_{\bf q}/k_BT_{\rm p}}-1)^{-1}$. Compared to the
normal state \cite{wellstood94}, we have inserted the normalized
density of states (DOS) $N(E)$. For a superconductor with energy gap
$\Delta(T)$, the DOS also depends on temperature
$N(E,T)=|E|/\sqrt{E^2-\Delta(T)^2}\Theta(E^2-\Delta(T)^2)$, where
$\Theta(x)$ is the Heaviside step function. Electrons emit energy to
phonons at the rate $P_e = N(E_F)\int dE_{{\bf k}} N(E_{{\bf
k}})f(E_{{\bf k}},T_{\rm e})\int d^3q D_p({\bf q})\epsilon_{\bf
q}\tau_{\bf k,k-q}^{-1}$. $D_p({\bf q})$ is the phonon DOS and
$N(E_F)$ is the normal-state DOS at the Fermi level. Writing a
similar expression for absorption of phonons $P_a$, we obtain the
net heat flux, $P_{\rm ep} = P_e-P_a$. Calculating the rate and the
matrix elements from the quasiclassical theory for clean
superconductors \cite{kopnin} we find the e-p heat flux
\begin{eqnarray}\label{epsc}
&&P_{\rm ep} = -\frac{\Sigma
\mathcal{V}}{96\zeta(5)k_B^5}\int_{-\infty}^{\infty} dE\, E
\int_{-\infty}^{\infty} d\epsilon \,\epsilon^2{\rm
sign}(\epsilon)M_{E,E+\epsilon} \nonumber \\
&&\times \big[\coth(\frac{\epsilon}{2k_BT_{\rm
p}})(f_E^{(1)}-f_{E+\epsilon}^{(1)})-f_E^{(1)}f_{E+\epsilon}^{(1)}+1\big].
\end{eqnarray}
Here, $f_E^{(1)}=f(-E)-f(E)$; for equilibrium, $f_E^{(1)}=
1-2f(E,T_{\rm e})=\tanh(\frac{E}{2k_BT_{\rm e}})$, and
$M_{E,E+\epsilon}$ is given by
$M_{E,E'}=N(E)N(E')[1-\frac{\Delta^2(T_{\rm e})}{EE'}]$.
In the regime $T_{\rm p} \ll T_{\rm e}\ll \Delta/k_B$ we obtain
$P_{\rm ep} \simeq \frac{64}{63\zeta(5)}\Sigma \mathcal{V} T_{\rm
e}^5e^{-\Delta/k_BT_{\rm e}}$,
which is by a factor $0.98e^{-\Delta/k_BT_{\rm e}}$ smaller than
the result for the normal state [Eq. \eqref{e-p-heat-norm} with
$T_{\rm p}\ll T_{\rm e}$].

Figure \ref{sample} shows a typical configuration of our
experiments. The samples were made by electron beam lithography and
shadow evaporation. The parameters of the structures are given in
Table \ref{table}. The aluminium block in the centre of Fig.
\ref{sample} is the volume in which energy relaxation is
investigated. Two small and two large tunnel junctions connect the
island to aluminium leads. The hot QP:s are injected via one of the
small tunnel junctions in series with a large one. Because of the
large asymmetry of junction parameters, essentially all the power is
injected by the small junction. The steady-state distribution on the
island is deduced from the current-voltage curves (IVs) of the
opposite pair of junctions. We observe the QP current of only the
small junction; the large junction remains in the supercurrent
state. Measurements in a configuration with two small junctions in
series as injectors and the two large junctions in series as probes
were also made with essentially identical results.

\begin{table}
\caption{Sample dimensions and junction resistances.} \label{table}
\begin{tabular}{|c|cc|ccc|}

\hline
 \multirow{1}{*}{Sample} & \multicolumn{1}{c}{volume ($\mu$m$^3$)} \vline & \multicolumn{1}{c}{$R_1,R_2,R_3,R_4$ (k$\Omega$)} \vline \\
\hline
A & 21  $\cdot$ 1.5  $\cdot$ 0.44 & 840, 4, 4, 1160 \\
B & 4.9  $\cdot$ 1.5  $\cdot$ 0.44 &  760, 5.7, 5.7, 1290 \\
C & 4.9  $\cdot$ 1.5  $\cdot$ 0.44 & 485, 20, 20, 980 \\
\hline
\end{tabular}
\end{table}

\begin{figure}
\begin{center}
\includegraphics[width=0.49\textwidth]{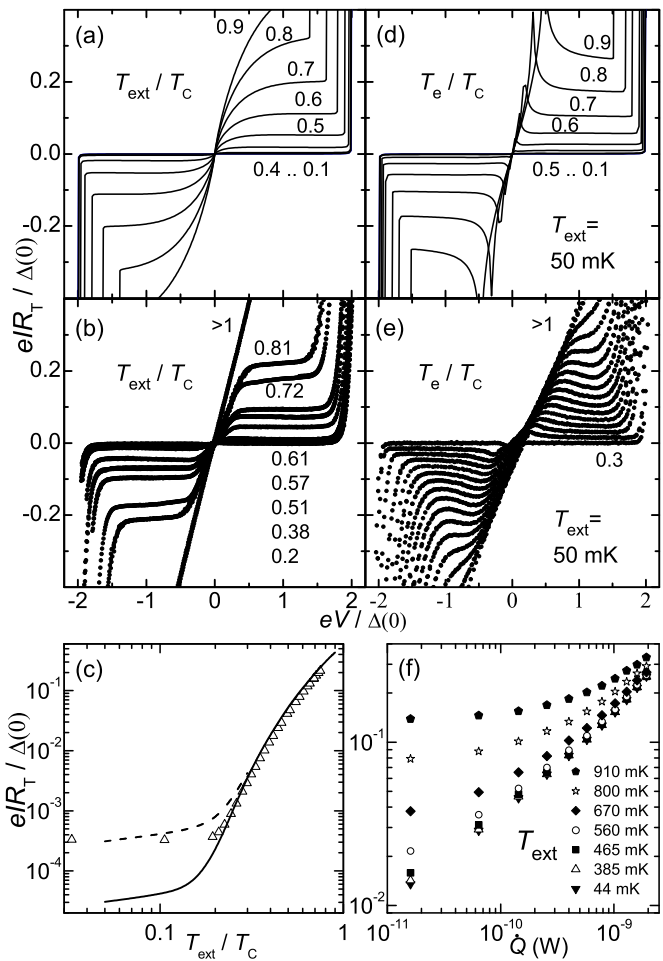}
\end{center}
\caption{Tunnel currents under equilibrium and quasi-equilibrium
conditions for a superconductor. Theoretical (a) and experimental
(b) IVs of a junction at several bath temperatures. (c) Theoretical
and experimental currents at $eV=\Delta$. The two theory lines
correspond to two different realistic pair breaking parameters
$\gamma = 10^{-3}$ (upper curve) and $\gamma = 10^{-4}$ (lower
curve). (d) Calculated IVs when the two leads of the junction have
different temperatures. (e) The measured IVs under a few injection
conditions. (f) The current in Sample A on the plateau between the
initial peak and the rise of the current at the conduction threshold
around $2\Delta/e$.
} \label{fig:IVs}
\end{figure}

If the island is at temperature $T_{\rm e}$ and the leads at $T_{\rm
ext}$, the QP current $I$ is given by $eR_TI = \int dE N(E-eV,T_{\rm
e})N(E,T_{\rm ext})[f(E-eV,T_{\rm e})-f(E,T_{\rm ext})]$. Here $R_T$
is the normal state resistance of the junction.  For equal island
and lead temperatures, $T_{\rm e} = T_{\rm ext}$, the calculated (a)
and measured (b) IVs for various $T_{\rm e}/T_C$ are shown in
Fig.~\ref{fig:IVs}. Wide plateaus in the regime $0 < eV < 2\Delta$
emerge due to the thermal QP current; its value at $eV =\Delta$ is
shown in (c). The agreement between experiment and theory is good
down to $T_{\rm e}/T_C \simeq 0.25$. To match the data to the theory
also at lower temperatures one can use the pair-breaking parameter
$\gamma\equiv \Gamma/\Delta$ which yields a smeared density of
states, $N(E,T)=|{\rm Re}
(E+i\Gamma)/\sqrt{(E+i\Gamma)^2-\Delta(T_{\rm e})^2}|$. In the
figure we show lines with $\gamma=10^{-4}$ and $\gamma=10^{-3}$.
Since at higher temperatures no fit parameter is needed, we focus
our analysis to the range $0.3 < T_{\rm e}/T_C < 1$ to be on the
safe side.

Figure \ref{fig:IVs} (d) shows the calculated IVs of the probe
junction, assuming that only the island temperature $T_{\rm e}$ is
elevated, and the leads remain at $T_{\rm ext}=0.05T_C$. This is the
expected behaviour under power injection, provided the e-e
relaxation is strong and that the junctions are opaque enough not to
conduct heat from the island into the leads. A peak in the IVs
arises at $eV=\Delta(T_{\rm ext})-\Delta(T_{\rm e})$. In Fig.
\ref{fig:IVs} (e) we show the corresponding measured curves at
various levels of injected power. The resemblance between (d) and
(e) is obvious but the features in experimental curves are broadened
in comparison to those from the theory, which is common for small
junctions \cite{steinbach01}. In the data analysis we next find the
minimum current in the plateau-like regime at bias voltages between
the "matching" peak and the strong onset of QP current. This current
is converted into temperature by comparing it to the $T_{\rm e}$
dependent minimum current of the theoretical IVs.

The power $\dot{Q}(V)$ deposited on the island by a biased junction
is given by
$e^2R_T\dot{Q}(V)= \int (E-eV)N(E-eV,T_{\rm e})N(E,T_{\rm
ext})[f(E,T_{\rm ext})-f(E-eV,T_{\rm e})]dE$.
This equation allows us to determine the injected power, as well as
the heat flux through all the junctions.
There are two features to note: (i) Since typical injection voltages
in the experiment are $V \gg \Delta/e$, it is sufficient to assume
that the power injected into the island equals $IV/2$, i.e., it is
divided evenly between the two sides of the junction: the junction
behaves essentially as a normal junction, where this statement is
true always. (ii) The heat conductance of the (probing) junction is
almost constant over a wide range of voltages within the gap region.
Its value is low and can be neglected under most experimental
conditions. Yet, to test this, we varied the resistances of the
large tunnel junctions by a factor of five between samples A and C,
without a significant effect on the results. Figure \ref{fig:IVs}
(f) shows the current on the plateau, as described in the previous
paragraph, as a function of power injected, at various bath
temperatures. In a wide range, from 30 mK up to 380 mK, the
behaviour is almost identical: only the higher temperature among
$T_{\rm e}$ and $T_{\rm p}$ plays a role, in consistence with the
theoretical discussion. Therefore we compare the experimental
results at the base phonon temperature (of about 50 mK) to the
theoretical results for $T_{\rm p}\ll T_{\rm e}$ in what follows.

We studied $P_{\rm ep}$ in the normal state as well by applying a
magnetic field of about 120 mT to suppress the superconductivity and
measuring the partial Coulomb blockade (CB) signal \cite{cbt94}.
Like in the superconducting state, two regimes are possible. In
equilibrium the results of Ref. \cite{cbt94} apply. Under injection,
the typical situation is such that $T_{\rm ext} \ll T_{\rm e}$,
which we discuss now in more detail. The tunnelling rates in a state
with an extra charge $n$ for adding ($+$) or removing ($-$) an
electron to the normal island with electrostatic energy change
$\Delta F^{\pm}(n)=\pm 2E_C(n\pm 1/2)\mp eV/2$ are
\begin{equation} \label{e1}
\Gamma^{\pm}(n)=\frac{1}{e^2R_T}\int_{-\infty}^\infty dE \,
f_1(E)[1-f_2(E-\Delta F^{\pm}(n))].
\end{equation}
Here $E_C=e^2/2C_\Sigma$ is the charging energy of the island with
the total capacitance $C_\Sigma$, and $f_1$ and $f_2$ are the
distributions on the source and target electrodes. For equilibrium
distribution $f_i(E)=(1+e^{E/k_BT_i})^{-1}$ with $T_1=T_2$, Eq.
\eqref{e1} yields the result of Ref.~\cite{cbt94}. Here we have the
opposite limit of low bath temperature, $T_{\rm ext}=T_1 \ll
T_2=T_{\rm e}$. For $T_1=0$, $f_1(E)=1-\Theta(E)$, yielding
$\Gamma^{\pm}(n)=(k_BT/e^2R_T)\ln(1+e^{-\Delta F^{\pm}(n)/k_BT_{\rm
e}})$. The current into the island is $I=e\sum_{n=-\infty}^{\infty}
\sigma(n)[\Gamma^+(n)-\Gamma^-(n)]$ where $\sigma(n)$ is the
probability of having $n$ extra electrons on the island. Since
$\sum_{n=-\infty}^{\infty} n\sigma(n)=0$ by symmetry, and
$\sum_{n=-\infty}^{\infty} \sigma(n)=1$, we find for the
differential conductance up to the first order in $E_C/k_BT_{\rm e}$
\begin{equation} \label{ecbt}
\frac{G^{\rm neq}}{G_T} = 1-\frac{E_C}{2k_BT_{\rm
e}}\frac{1}{\cosh^2(eV/4k_BT_{\rm e})}.
\end{equation}
The depth of the conductance minimum at $V=0$ is $ \Delta G/G_T =
E_C/2k_BT_{\rm e}$ which is 50\% larger than that in the equal
temperature case. To find the width at half minimum we need to solve
$\cosh^2(eV_\pm/4k_BT_{\rm e})=2$.
The full width is $V_{1/2}^{\rm neq}=|V_+-V_-|$ or $ V_{1/2}^{\rm
neq} = 4\ln(3+2\sqrt{2})k_BT_{\rm e}/e$. This is about 65\% of the
equal-temperature value, $V_{1/2}^{\rm eq}\simeq 10.88k_BT_{\rm
e}/e$ \cite{cbt94}.

\begin{figure}
\begin{center}
\includegraphics[width=0.5\textwidth]{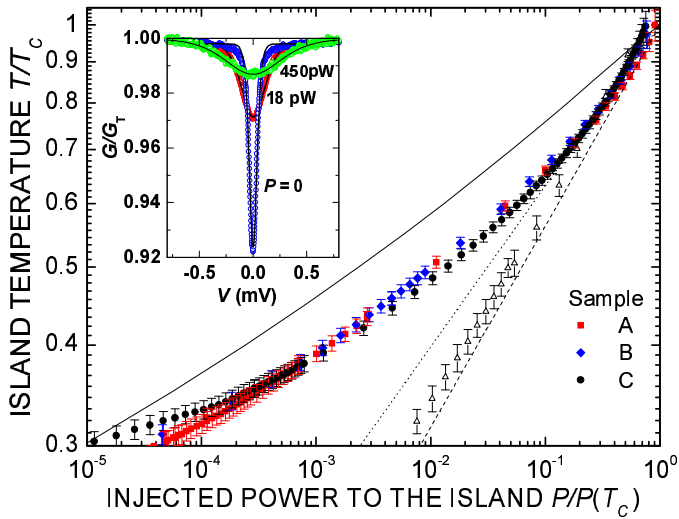}
\end{center}
\caption{(color online) Energy relaxation from theory and
experiment. The data in the superconducting state are from Sample
A (squares), B (diamonds), and C (circles). The open triangles are
from Sample C in the normal state. The solid line is the result of
Eq. \eqref{epsc} in the superconducting state. The dotted line
indicates $P/P(T_C)=(T/T_C)^5$, and the dashed line
$P/P(T_C)=(T/T_C)^4$. The inset shows three Coulomb peaks measured
in the normal state under different levels of power injection: the
solid lines are theoretical fits to them.} \label{fig:theoryvsexp}
\end{figure}

Figure \ref{fig:theoryvsexp} is a collection of the data at the base
temperature ($\simeq 50$ mK), in form of island temperature $T_{\rm
e}/T_C$ as a function of injected power. The superconducting state
was measured for the three samples. The power has been normalized by
that at $T_C$, to present data from different samples on the same
footing. For samples A, B and C, $P(T_C)= 14$ nW, $3$ nW, and $3$
nW, respectively. The data on the three samples are mutually
consistent. The quasiclassical result for a superconductor is shown
by a solid line. The normal state data were taken for Sample C which
is ideal for a measurement of the island temperature via partial CB:
it has $E_C/k_B\simeq 20$ mK yielding an approximately $8$\% deep
Coulomb dip in conductance at zero bias at 90 mK (see the inset of
Fig. \ref{fig:theoryvsexp}). The two large junctions were used for
probing and the small ones for power injection. We first checked
that the value $V_{1/2}^{\rm eq}$ yields a good quantitative
agreement with the equilibrium temperature data over the whole range
of the experiment. Next we measured the quasi-equilibrium electronic
temperature under injection. The low base temperature permits the
use of the expression of $V_{1/2}^{\rm neq}$ above to extract
$T_{\rm e}$ in the range displayed in Fig. \ref{fig:theoryvsexp}.
Power law type behavior can be observed over the whole temperature
range $0.3T_C < T_{\rm e} \lesssim T_C$. The data approach those of
the superconducting state near $T_C\simeq 1.45$ K, as expected. The
power law for $P_{\rm ep}$ is, however, better approximated by
$T_{\rm e}^4$ (dashed line) instead of $T_{\rm e}^5$ (dotted line)
of Eq. \eqref{e1}, yielding a deviation of the same sign with
respect to the basic theories as in the superconducting state.

The experimental data demonstrate that e-p coupling in a
superconductor is weaker than in the normal state, by two orders of
magnitude at $T_{\rm e}/T_C = 0.3$. But, like in the relaxation time
experiments in a superconductor \cite{day03,kozorezov01,barends08},
the energy flux is larger than that from the quasiclassical theory
\cite{kaplan76,kopnin}. This observation could suggest that the
electron relaxation rate both in the superconducting and in the
normal state might be sensitive to the microscopic quality and the
impurity content of the particular film \cite{sergeev00}. The
impurity effects on the e-p relaxation are controlled by the
parameter $q\ell$ where $\ell$ is the electronic mean free path and
$q=k_BT_{\rm e}/\hbar u$ is the wave vector of an emitted phonon
with energy of the order of the electronic temperature. With the
speed of sound $u\sim 5000$ m/s and $\ell \sim 20$ nm in our
samples, we have $q\ell\sim 0.5 ~{\rm K}^{-1}~T_{\rm e}$. Thus the
impurity effects can become essential below 1 K. More theoretical
studies are thus needed of the impurity effects on the e-p
interaction in the superconducting state.

Our experiments were performed on three samples with very different
lengths and junction parameters but they yielded essentially
identical results when normalized by the island volume. Therefore we
believe that issues like thermal gradients, slow electron-electron
relaxation, and the presence of tunnel contacts have only a minor
influence on the results. The data thus yield the intrinsic energy
relaxation of QP:s in the superconducting and in the normal state.
In summary, the experiment follows qualitatively the theoretical
model that we presented. On the quantitative level there is a
substantial discrepancy especially for superconductors, which would
imply that one needs to invoke an extra relaxation channel to
account for. Solution of this quantitative disagreement and
experiments at still lower temperatures remain as topics of future
work.

We thank M. Gershenson, H. Courtois, F. Hekking, A. Niskanen, T.
Heikkil\"a, and Yu. Galperin for useful discussions. This work was
supported by the NanoSciERA project "NanoFridge", by Russian
Foundation for Basic Research grant 06-02-16002, and by the Academy
of Finland.

\end{document}